 \documentclass[final,authoryear,1p,times,twocolumn]{elsarticle}
 \usepackage[final]{pdfpages}

\usepackage{amssymb}

 \usepackage{lineno}

\journal{J. of Atmospheric and Solar-Terrestrial Physics}

\begin{document}

\begin{frontmatter}




\title{Testing an astronomically-based decadal-scale empirical harmonic climate model versus the IPCC (2007) general circulation climate models}



\author{Nicola Scafetta $^{1}$}

 \address{$^{1}$ACRIM (Active Cavity Radiometer Solar Irradiance Monitor Lab)  \& Duke University, Durham, NC 27708, USA.}

\begin{abstract}
We compare the performance of a recently proposed empirical climate model based on astronomical harmonics  against all  CMIP3 available general circulation  climate models (GCM)  used by the IPCC (2007) to interpret the 20$^{th}$ century global surface temperature. The proposed astronomical empirical climate model assumes that the climate is resonating with, or synchronized to a set of natural harmonics that, in previous works (Scafetta, 2010b, 2011b), have been associated to the solar system planetary motion, which is mostly determined by Jupiter and Saturn.  We show that the  GCMs fail to  reproduce  the major decadal and multidecadal oscillations found in the global surface temperature record from 1850 to 2011.   On the contrary, the proposed harmonic model (which herein uses cycles with 9.1, 10-10.5, 20-21,60-62 year periods) is found to well reconstruct the observed climate oscillations from 1850 to 2011, and it is able to forecast the climate oscillations from 1950 to 2011 using the data covering the period 1850-1950, and vice versa.  The  9.1-year cycle is shown  to be likely related to a decadal Soli/Lunar tidal oscillation, while the  10-10.5, 20-21 and 60-62 year cycles are synchronous to solar and heliospheric planetary oscillations. We show that the IPCC GCM's claim that all warming observed from 1970 to 2000 has been anthropogenically induced is  erroneous  because of the  GCM failure in reconstructing the quasi 20-year and 60-year climatic cycles. Finally, we show how the presence of these large natural cycles can be used to correct the IPCC projected anthropogenic warming trend for the 21$^{st}$ century. By combining this corrected trend with the natural cycles, we show that the temperature may not significantly increase during the next 30 years mostly because of the negative phase of the 60-year cycle. If multisecular natural cycles (which according to some authors have significantly contributed to the observed 1700-2010 warming  and may contribute to an additional natural cooling by 2100) are ignored, the same IPCC projected anthropogenic emissions would imply a global  warming by about 0.3-1.2 $^oC$ by 2100, contrary to the IPCC 1.0-3.6 $^oC$ projected warming. The results of this paper reinforce  previous claims that the relevant physical mechanisms that explain the detected climatic  cycles are still missing in the current GCMs and that climate variations at the  multidecadal scales  are astronomically induced  and, in first approximation, can be forecast.
\end{abstract}


\end{frontmatter}


\section{Introduction}

  Herein,  we test the performance of a recently proposed astronomical-based empirical harmonic climate model   \citep{scafettanew,scafett2011b} against all general circulation climate models (GCMs) adopted by the   \cite{IPCC} to interpret climate change during the last century. A large Supplement file with all GCM simulations herein studied plus additional information is added to this manuscript. A reader is invited to look at the figures depicting the single GCM runs there reported to have a feeling about the performance of these models.

  The astronomical harmonic model assumes that the climate system is resonating with or  is synchronized to a set of natural frequencies of the  solar system.
  The synchronicity between solar system oscillations and climate cycles has been extensively discussed and argued in  Scafetta (2010a, 2010b, 2011b), and in the numerous references cited in those papers. We used the velocity of the Sun relative to the barycenter of the solar system and a record of historical mid-latitude aurora events. It was observed that  there is a good synchrony of frequency and phase between multiple astronomical cycles with periods between 5 to 100 years and equivalent cycles found in the climate system. We refer to those works for details and statistical tests. The major hypothesized mechanism is that the planets, in particular Jupiter and Saturn, induce solar or heliospheric oscillations that induce equivalent oscillations in the electromagnetic  properties of the upper atmosphere. The latter induces similar cycles in the cloud cover and in the terrestrial albedo forcing the climate to oscillate in the same way. The soli/lunar tidal cyclical dynamics also appears to play an important role  in climate change at specific frequencies.

This work focuses only on the major decadal and multidecadal oscillations of the climate system, as observed in the global surface temperature data since AD 1850. A more detailed discussion about the interpretation of the secular climate warming trending since AD 1600 can be found in \cite{Scafetta2007} and in \cite{Scafetta3} and in numerous other references there cited. About the millenarian cycle since the Middle Age a discussion is present in \cite{Scafetta55} where the relative contribution of solar, volcano and anthropogenic forcing is also addressed, and in the numerous references cited in the above three papers. Also correlation studies between the secular trend of the temperature and the geomagnetic aa-index, the sunspot number and the solar cycle length address the above issue and are quite numerous: for example \citep{Hoyt,Sonnemann,Thejll}. Thus, a reader interested in better understanding the secular climate trending topic is invited to read those papers. In particular, about the 0.8$^oC$ warming trending observed since 1900 numerous empirical studies based on the comparison between the past climate secular and multisecular patterns and equivalent solar activity patterns have concluded that at least 50-70\% of the observed 20$^{th}$ century  warming could be associated to the increase of solar activity observed since the Maunder  minimum of the 17$^{th}$ century: for example see \citep{Scafetta2007,Scafetta3,Loehle,Soon2009,Soon2011,Kirkby,Hoyt,Courtillot44,Thejll,Weihong,Eichler}. Moreover, \cite{Humlum} noted that the natural multi-secular/milennial climate cycles observed during the  late Holocene climate change clearly suggest that the secular 20$^{th}$ century warming could be mostly due to these longer natural cycles, which are also expected to cool the climate during  the 21$^{th}$ century. A similar conclusion has been reached by another study focusing on the multi-secular and millennial cycles observed in the temperature  in the central-eastern Tibetan Plateau during the past 2485 years \citep{Liu}. For the benefit of the reader, in section 7 in the Supplement file the results reported in two of the above papers are very briefly presented to graphically support the above claims.

It is important to note that  the above empirical results contrast greatly with the GCM estimates adopted by the IPCC  claiming that  more than 90\% of the warming observed since 1900 has been anthropogenically induced (compare figures 9.5a and 9.5b  in the IPCC report which are reproduced in section 4 in the Supplement File). In the above papers it has been often argued that the current GCMs miss important climate mechanisms such as, for example, a modulation of the cloud system via a solar induced modulation of the cosmic ray incoming flux, which would greatly amplify the climate sensitivity to  solar changes by modulating the terrestrial albedo \citep{scafett2011b,Kirkby,Svensmark98,Svensmark,Shaviv}.

In addition to a well-known decadal climate cycle commonly associated to the Schawbe solar cycle by numerous authors \citep{Hoyt}, several studies  have emphasized  that the climate system is characterized by a quasi bi-decadal (from 18 yr to 22 yr) oscillation and by a quasi 60-year oscillation \citep{Stockton,Currie,Cook,Agnihotri,Klyashtorin,Sinha,Yadava,Jevrejeva,Knudsen,Davis,scafettanew,Weihong,Mazzarella,scafett2011b}.
    For example, quasi 20-year and 60-year large cycles are clearly detected in all global surface temperature instrumental records of both hemispheres since 1850 as well as in numerous astronomical records.    There is a phase synchronization between these terrestrial and astronomical cycles. As argued in \cite{scafettanew}, the observed quasi bidecadal climate cycle may also be around a 21-year periodicity because of the presence of the 22-year solar Hale magnetic cycle, and there may also be an additional influence of the 18.6-year soli/lunar nodal cycle. However, for the purpose of the present paper, we can ignore these corrections which may require other cycles at 18.6 and 22 years. In the same way, we ignore other possible slight cycle corrections due to the interference/resonance with other planetary tidal cycles and with the 11-year and 22-year solar cycles, which are left to another study.

    About the 60-year cycle it is easy to observe that the global surface temperature experienced major maxima in 1880-1881, 1940-1941 and 2000-2001. These periods occurred during the Jupiter/Saturn great conjunctions when the two planets were quite close to the Sun and the Earth. This events occur every three J/S synodic cycles. Other local temperature maxima occurred during the other J/S conjunctions, which occur every about 20 years: see figures 10 and 11 in \cite{scafettanew}, where this correspondence is shown in details through multiple filtering of the data. Moreover, the tides produced by Jupiter and Saturn in the heliosphere and in the Sun have a period of about $0.5/(1/11.86-1/29.45)\approx10$ years plus the 11.86-year Jupiter orbital tidal cycles. The two tides beat generating an additional cycle  at about $1/(2/19.86-1/11.86)=61$ years \citep{scafett2011b}. Indeed, a quasi 60-year climatic oscillations have likely an astronomical origin because the same cycles are found in numerous secular and millennial aurora and other solar related records   \citep{Charvatova,Komitov,Ogurtsov,Patterson,Yu,Scafetta55,scafettanew,Mazzarella,scafett2011b}.

A 60-year cycle is even referenced in ancient Sanskrit texts among the observed monsoon rainfall cycles \citep{Iyengar}, a fact confirmed by modern monsoon studies \citep{Agnihotri}. It is also observed in the sea level rise since 1700 \citep{Jevrejeva} and in numerous ocean and terrestrial records for centuries \citep{Klyashtorin}. A natural 60-year climatic cycle associated to planetary astronomical cycles may also explain the origin of 60-year cyclical calendars adopted in traditional Chinese, Tamil and Tibetan civilizations \citep{Aslaksen}. Indeed, all major ancient civilizations knew about the 20-year and 60-year astronomical cycles associated to Jupiter and Saturn \citep{Temple}.

In general,  power spectrum evaluations have shown that frequency peaks with periods of about 9.1, 10-10.5, 20-22  and 60-63 years  are the most significant ones  and are common between astronomical and climatic records \citep{scafettanew,scafett2011b}. Evidently, if climate is described by a set of harmonics, it can be in first approximation reconstructed and forecast by using a  planetary harmonic constituent analysis methodology similar to the one that was first proposed by Lord Kelvin \citep{Kelvin,scafett2011b} to accurately reconstruct and predict tidal dynamics.   The harmonic constituent model is just a superposition of several harmonic terms of the type

\begin{equation}\label{eqa0}
    F(t)=A_0 + \sum_{i=1}^N A_i\cos(\omega_i t + \phi_i).
\end{equation}
whose frequencies $\omega_i$ are deduced from the astronomical theories and the  amplitude $A_i$ and phase $\phi_i$ of each  harmonic  constituent are empirically determined using regression on the available data, and then the model is used to make forecasts. Several harmonics are required: for example, most locations in the United States use computerized forms of Kelvin's tide-predicting machine with 35-40 harmonic constituents for predicting local tidal amplitudes \citep{Ehret}, so a reader should not be alarmed if many harmonic constituents may be needed to accurately reconstruct  the climate system.

Herein we show that a similar harmonic empirical methodology can, in first approximation, reconstruct and forecast global climate changes at least on a decadal and multidecadal scales, and  that  this methodology works much better than the current GCMs adopted by the IPCC in 2007. In fact, we will show that the IPCC GCMs fail to reproduce the observed climatic oscillations at multiple temporal scales. Thus, the computer models adopted by the IPCC in 2007 are found to be missing the important physical mechanisms responsible for  the major observed climatic oscillations. An important consequence of this finding is that these GCMs have seriously misinterpreted the reality by significantly overestimating the anthropogenic contribution,  as also other authors have recently claimed \citep{Douglass,Lindzen,Spencer}. Consequently, the IPCC projections for the $21^{st}$ century should not be trusted.

\section{The IPCC GCMs do not reproduce the global surface temperature decadal and multidecadal cycles}

Figure 1 depicts the monthly global surface temperature anomaly (from the base period 1961-90) of the Climatic Research Unit  (HadCRUT3) \citep{Brohan} from 1850 to 2011 against an advanced general circulation model average simulation \citep{Hansen}, which has been slightly shifted downward for visual convenience. The chosen units are the degree Celsius in agreement with the climate change literature referring to temperature anomalies.
The GISS ModelE is one of the major GCMs adopted by the IPCC \citep{IPCC}.  Here we study all available climate model simulations for the 20th century collected by Program for Climate Model Diagnosis and Intercomparison (PCMDI) mostly during the years 2005 and 2006, and this archived data constitutes phase 3 of the Coupled Model Intercomparison Project (CMIP3). These GCMs use  the observed radiative forcings (simulations ``tas:20c3m'')  adopted by the IPCC (2007). All GCM simulations are depicted and analyzed in Section 2 of the Supplement file added to this paper. These GCM simulations cover a period that may begin during the second half of the 19$^{th}$ century and end during the 21$^{th}$ century. The following calculations are based on the maximum overlapping period between each model simulation and the 1850-2011 temperature period.
The  CMIP3 GCM simulations analyzed here  can be downloaded from Climate Explorer web-site: see the Supplement file for details.

\begin{figure}
\includegraphics[angle=0,width=30pc]{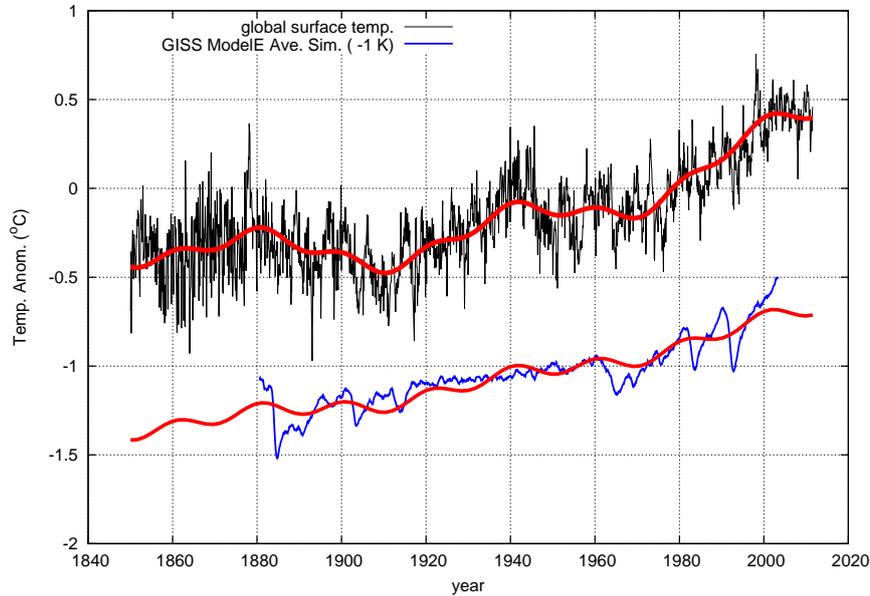}
 \caption{
 Global surface temperature (top (http://www.cru.uea.ac.uk/cru/data/temperature/) and GISS ModelE average simulation (bottom). The records are fit with Eq 5. Note also the large volcano eruption signatures that appear clearly overestimated in the GCM's simulation.
 }
\end{figure}

A simple visual inspection suggests that the temperature presents a quasi 60-year cyclical modulation oscillating around an upward trend \citep{scafettanew,Loehle}. In fact, we have the following 30-year trending patterns: 1850-1880, warming; 1880-1910, cooling; 1910-1940, warming; 1940-1970, cooling; 1970-2000, warming; and it is almost steady or presents a slight cooling since 2001 (2001-2011.5 rate = -0.46 $\pm0.3 ~^oC/century$). Other global temperature reconstructions, such as the GISSTEM \citep{Hansen} and the GHCN-Mv3 by NOAA, present similar patterns (see Section 1 in the Supplement file). Note that GISSTEM/1200 presents a slight warming since 2001 (2001-2011.5 rate = +0.47 $\pm0.3 ~^oC/century$), which appears to be due to the GISS poorer temperature sampling during the last decade of the Antarctic and Arctic regions that were artificially filled with a  questionable 1200 km smoothing methodology \citep{Tisdale}. However, when a 250 km smooth methodology is applied, as in GISSTEM/250, the record shows a slight cooling during the same period (2001-2011.5 rate = -0.16 $\pm0.3 ~^oC/century$).   HadCRUT data has much better coverage of the Arctic and Southern Oceans that GISSTEM and, therefore, it is likely more accurate. Note that CRU has recently produced an update of their SST ocean record, HadSST3, \citep{Kennedy}, but it stops in 2006 and  was not merged yet with the land record. This new corrected record presents an even clearer 60-year modulation than the HadSST2 record because in it the slight cooling from 1940 to 1970 is clearer \citep{Mazzarella}.

Indeed, the 60-year cyclicity with peaks in 1940 and 2000 appears quite more clearly in numerous regional surface temperature reconstructions that show a smaller secular warming trending. For example, in  the United States \citep{Daleo}, in the  Arctic region \citep{Soon2009}, in several single stations in Europe and other places \citep{Courtillot44} and in China \citep{Soon2011}. In any case, a 60-year cyclical modulation is present for both the Norther and Southern Hemisphere and for both Land and Ocean regions \citep{scafettanew} even if it may be partially hidden by the upward warming trending.  The 60-year modulation appears well correlated to a recently proposed solar activity reconstruction \citep{Loehle}.

The 60-year cyclical modulation of the temperature from 1850 to 2011 is further shown in Figure 2 where the autocorrelation functions of the global surface temperature and of the GISS ModelE average simulation are compared. The autocorrelation function is defined as:

\begin{equation}\label{eq5652}
    r(\tau)=\frac{\sum_{t=1}^{N-\tau}(T_t-\bar{T}) (T_{t+\tau}-\bar{T})}{\sqrt{[\sum_{t=1}^{N-\tau}(T_t-\bar{T})^2 \sum_{t=\tau}^{N}(T_t-\bar{T})^2]}},
\end{equation}
where $\bar{T}$  is the average of the  N-data long temperature record and $\tau$ is the time-lag.
The autocorrelation function of the global surface temperature (Fig. 2A) and of the same record detrended of its quadratic trend (Fig. 2B) reveals the presence of a clear cyclical pattern with minima at about 30-year lag  and 90-year lag, and  maxima at about 0-year lag and 60-year lag. This pattern  indicates the presence of a quasi 60-year cyclical modulation in the record. Moreover, because both figures show the same pattern it is demonstrated  that the quadratic trend does not artificially creates the 60-year cyclicity.  On the contrary, the GISS ModelE average simulation produces a very different autocorrelation pattern lacking any cyclical modulation.  Figure 2C shows the autocorrelation function of the two records detrended also of their 60-year cyclical fit, and the climatic record appears to be characterized by a quasi 20-year smaller cycle, as deduced by the small but visible quasi regular 20-year waves, at least up to a time-lag of 70 years after which other faster oscillations with a decadal scale dominate the pattern. On the contrary, the autocorrelation function of the GCM misses both the decadal and bi-decadal oscillations and again shows a strong 80-year lag peak, absent in the temperature. The latter peak is due to the quasi 80-year lag between the two computer large volcano eruption signatures of Krakatoa (1883) and  Agung (1963-64),  and to the quasi 80-year lag between the  volcano signatures of Santa Maria (1902) and  El Chich\'{o}n (1982). Because this 80-year lag autocorrelation peak is not evident in the autocorrelation function of the global temperature we can conjecture that the GISS ModelE is significantly overestimating the volcano signature, in addition to not reproducing the natural decadal and multidecadal temperature cycles: this claim is further supported in Section 5 of the Supplement file.

\begin{figure}
\includegraphics[angle=0,width=30pc]{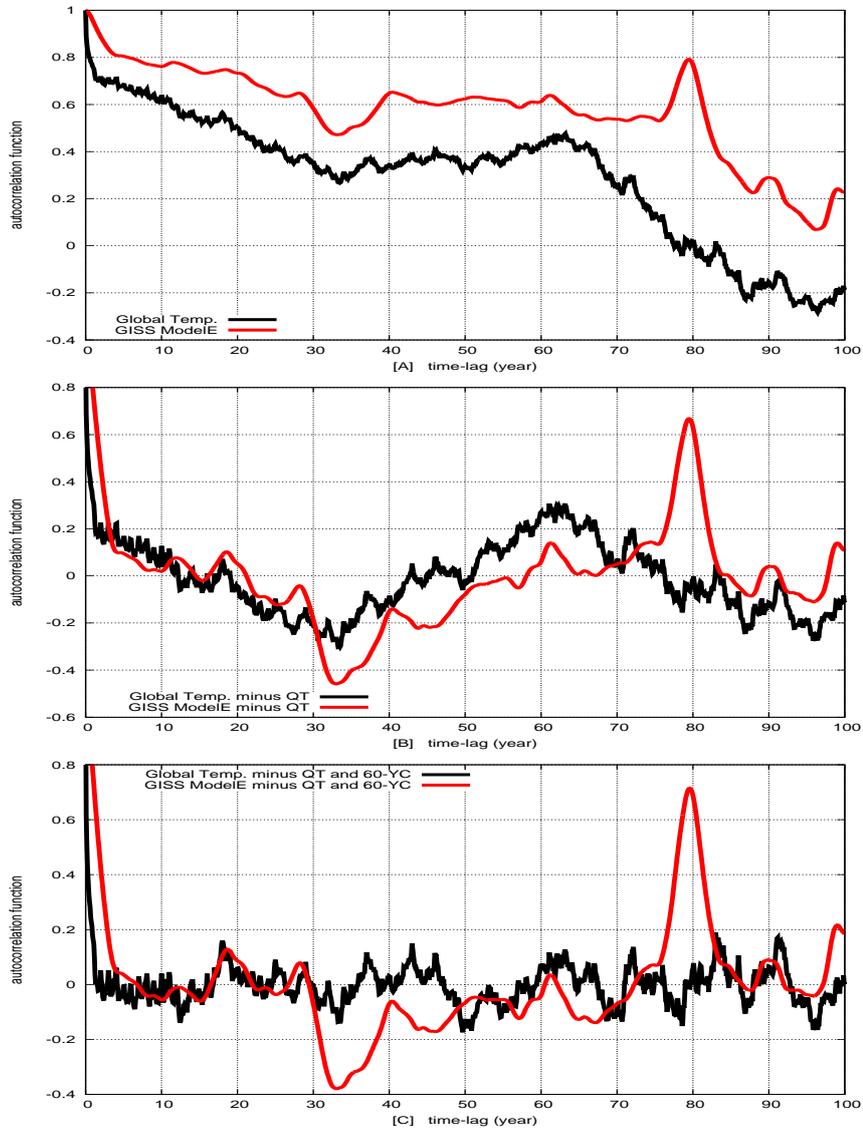}
 \caption{
 Autocorrelation function (Eq. 2) of the global surface temperature and of the GISS ModelE average simulation: [A] Original data; [B] data detrended of their quadratic fit; [C] The 60-year modulation is further detrended. Note the 60-year cyclical modulation of the autocorrelation of the temperature with minima at 30-year and 90-year lags and maxima at 0-year and 60-year lags, which is not reproduced   by the GCM simulation. Moreover, the computer simulation presents an autocorrelation peak at 80-year lag related to a pattern produced by volcano eruptions, which is absent  in the temperature.  See Section 5 in the supplement file for further evidences about the GISS ModelE serious overestimation of the volcano signal in the global surface temperature record.
 }
\end{figure}

A similar qualitative conclusion applies also to all other GCMs used by the IPCC, as shown in Section 2 of the Supplement file. The single GCM runs as well as their average reconstructions appear quite different from each other: some of them are quite flat until 1970, others are simply monotonically increasing. Volcano signals often appear overestimated. Finally, although   these GCM simulations present some kind of red-noise variability supposed to simulate the multi-annual, decadal and multi-decadal natural variability, a simple visual comparison among the simulations and the temperature record gives a clear impression that the simulated variability has nothing to do with the observed temperature dynamics. In conclusion, a simple visual analysis of the records suggests that  the temperature is characterized  10-year, 20-year and 60-year  oscillations that are simply not reproduced by the GCMs. This is also implicitly indicated by the very smooth and monotonically increasing pattern of their average reconstruction depicted in the IPCC figure SPM.5 (see Section 4 in the Supplement file).

 Figures 3A and 3B shows two power spectra estimates of the temperature records based on the Maximum Entropy Method (MEM) and the Lomb periodogram \citep{Press}. Four major peaks are found at periods of about 9.1, 10-10.5, 20-21 and 60-62 years: other common peaks are found but not discussed here. Both techniques produce the same spectra. To verify whether the detected major cycles are physically relevant and not produced by some unspecified noise or by the specific sequences, mathematical algorithms and  physical assumptions used to produce the HadCRUT record,  we have compared the same double power spectrum analysis applied to the three available global surface temperature records   (HadCRUT3, GISSTEM/250 and GHCN-Mv3) during their common overlapping time period (1880-2011): see also section 1 in the Supplement file. As shown in the figures the  temperature sequences present almost identical power spectra with major common peaks at about 9.1, 10-10.5, 20 and 60 years. Note that in \cite{scafettanew}, the relevant frequency peaks of the temperature were determined by comparing the power spectra of HadCRUT temperature records referring to different regions of the Earth such as those referring to the Northern and Southern hemispheres, and to the Land and the Ocean.  So, independent major global surface temperature records present the same major periodicities: a fact that further argues for the physical global character of the detected spectral peaks.

\begin{figure}
\begin{center}
\includegraphics[angle=0,width=30pc]{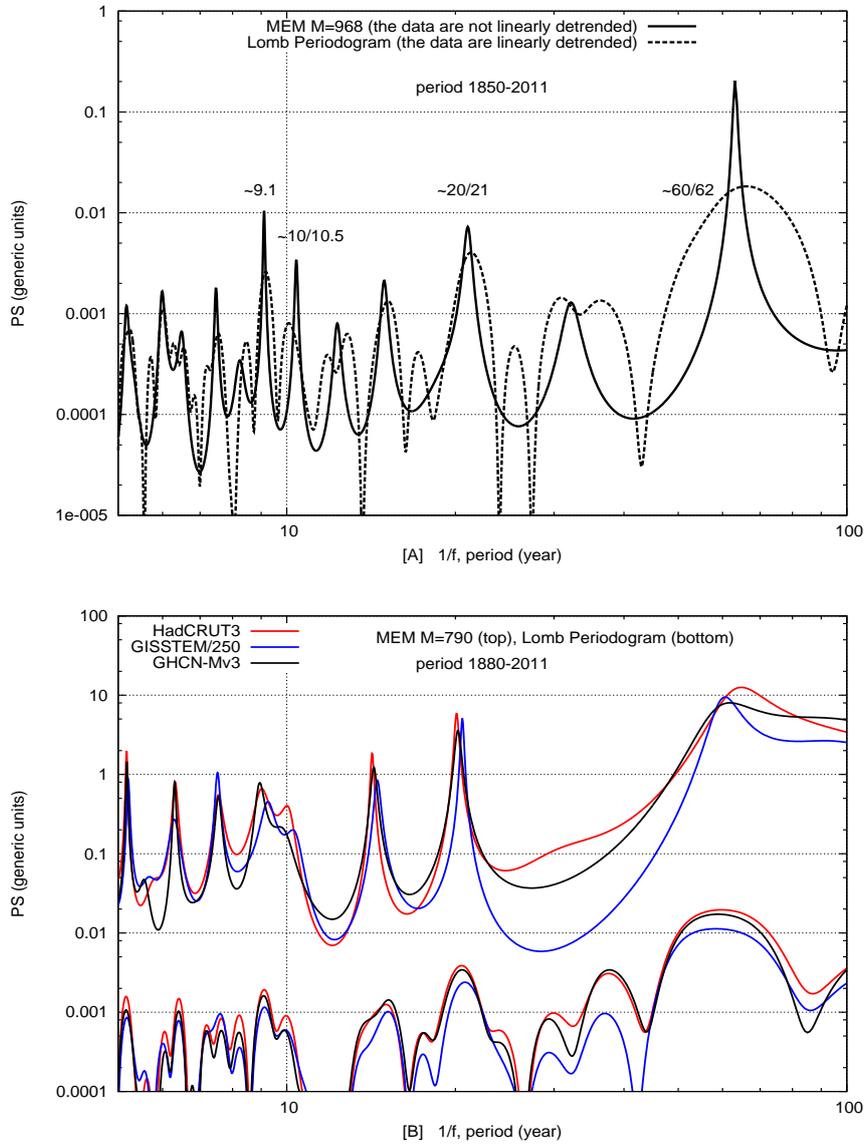}
\end{center}
 \caption{[A] Maximum Entropy Method (MEM) with M=N/2 (solid)  and the Lomb Periodogram (dash) of the HadCRUT3 global surface temperature monthly sampled from 1850 to 2011 (see Section 3 of the Supplement file for details and explanations). The two techniques produce the same peaks, but MEM produces much sharper peaks. The major four peaks are highlighted in the figure.  [B] As above for the HadCRUT, GISSTEM/250 and GHCN-Mv3 global surface temperature records during the period 1880-2011: see section 1 in the Supplement file. Note that the spectra are quite similar, but for GISSTEM the cycles are somehow slightly smoother and smallers than for the other two sequences, as the bottom curves show.  The result shows that all GCMs significantly fail in reproducing the 20-year and 60-year cycle amplitudes observed in the temperature record by an average factor of 3. }
\end{figure}

 Note that a methodology based on a spectral comparison of independent records is likely more physically appropriate than using purely statistical methodologies based on Monte Carlo randomization of the data, that may likely interfere with weak dynamical cycles.
Note also that a major advantage of MEM is that it produces much sharper peaks that allow a more detailed analysis of the \emph{low-frequency band} of the spectrum.  Section 5 in the Supplement file contains a detailed explanation about the number of poles M needed to let MEM to resolve the very-low frequency range of the spectrum: see also \cite{Courtillot1977}.

Because the temperature record presents major frequency peaks at about 20-year and 60-year periodicities plus an apparently accelerating upward trend, it is legitimate to extract these multidecadal patterns by fitting the  temperature record (monthly sampled) from 1850 to 2011 with the 20 and 60-year cycles plus a quadratic polynomial trend. Thus, we use a function $f(t)+p(t)$ where the harmonic component is given by

\begin{equation}\label{eq1}
    f(t)=C_1\cos\left[\frac{2\pi (t-T_1)}{60}\right]+C_2\cos\left[\frac{2\pi (t-T_2)}{20}\right],
\end{equation}
and the upward quadratic trending is given by

\begin{equation}\label{eqpt}
   p(t)=P_2*(t-1850)^2+P_1*(t-1850)+P_0~.
\end{equation}
 The regression values for the harmonic component are: $C_1=0.10\pm0.01~^oC$ and $C_2=0.040\pm0.005~^oC$, and the two dates are $T_1=2000.8 \pm0.5~AD$ and $T_2=2000.8 \pm 0.5~AD$. For the quadratic component we find: $P_0=-0.30\pm0.2~^oC$, $P_1=-0.0035\pm0.0005~^oC/yr$ and $P_2=0.000049\pm0.000002~^oC/yr^2$. Note that the two cosine phases are free parameters and the regression model gives the same phases for both harmonics, which  suggests that they are related. Indeed, this common phase date approximately coincides with the closest (to the sun) conjunction between Jupiter and Saturn, which occurred (relative to the Sun) on June/23/2000 ($\approx2000.5$), as better shown in \cite{scafettanew}.

It is important to stress that the above quadratic function $p(t)$ is just a convenient geometrical  representation of the  observed warming accelerating trend during the last 160 years, not outside the fitting interval. Another possible choice, which uses two linear approximations during the periods 1850-1950 and 1950-2011, has also be proposed \citep{Loehle}. However, our quadratic fitting trending cannot be used for forecasting purpose, and  it is not a component  of the astronomical harmonic model. Section 4 will address the forecast problem in details.

It is possible to test how well the IPCC GCM simulations reproduce the 20 and 60-year temperature cycles plus the upward trend from 1850 to 2011 by fitting their simulations with the following equation

\begin{equation}\label{eq2}
    m(t) =a*0.10\cos\left[\frac{2\pi (t-2000.8)}{60}\right]+ b*0.040\cos\left[\frac{2\pi (t-2000.8)}{20}\right] +c*p(t) +d~,
\end{equation}
where $a$, $b$, $c$ and $d$ are  regression coefficients. Values of $a$, $b$ and $c$ statistically compatible with the number 1 indicate that the model well reproduces the observed temperature 20 and 60-year cycles, and the observed upward temperature trend from 1850 to 2011. On the contrary, values of $a$, $b$ and $c$ statistically incompatible with 1  indicate that the model does not reproduce the observed temperature patterns.

\begin{figure}
\includegraphics[angle=0,width=30pc]{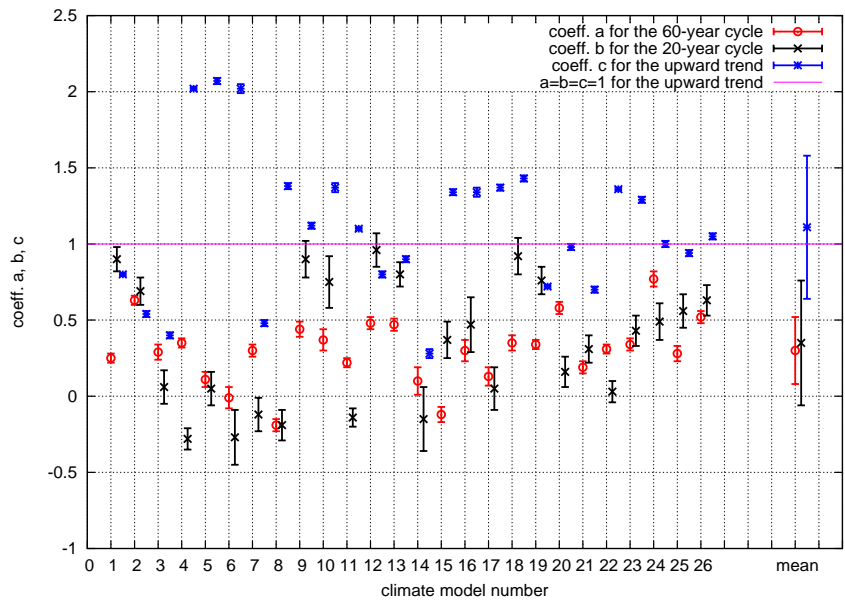}
 \caption{
Values of the regression coefficients $a$, $b$ and $c$ relative to the amplitude of the 60 and 20-year cycles, and the upward trend obtained by regression fit of the 26 GCM simulations of the 20$^{th}$ century used by the IPCC. See Table 1 and the Section 2 in the Supplement file for details.
 }
\end{figure}

The regression values for all GCM simulations are reported in Table 1.
 Figure 4 shows the values of the regression coefficients $a$, $b$ and $c$ for the 26 climate model  ensemble-mean records and all fail to well reconstruct both the 20 and the 60-year oscillations found in the climate record. In fact, the values of the regression coefficients $a$ and $b$ are always well below the optimum value of 1, and for some model these values are even negative. The average among the 26 models is $a=0.30\pm0.22$ and $b=0.35\pm0.42$, which are statistically different from 1. This result would not change if all available single GCM runs  are analyzed separately, as extensively shown in Section 2 of the Supplement file.

 About the capability of the GCMs of reproducing the upward temperature trend from 1850 to 2011, which is estimated by the regression coefficient $c$, we find a wide range of results. The average is $c=1.11\pm0.50$, which is centered close to the optimum value 1. This result explains why the multi-model global surface average simulation depicted in the IPCC figures 9.5 and SPM.5 apparently reproduces the 0.8 $^oC$ warming observed since 1900. However, the results about the regression coefficient $c$ vary greatly from model to model: a fact that indicates that these GCMs usually also fail to properly reproduce the observed upward warming trend from 1850 to 2011.

 Table 1 and the tables in Section 2 in the Supplement file also report the estimated reduced $\chi^2$ values between the measured GCM coefficients $a_m$, $b_m$ and $c_m$ (index ``m'' for model) and the values of the same coefficients $a_T$, $b_T$ and $c_T$ (index ``T'' for temperature) estimated for the temperature. The reduced $\chi^2$ (chi square) values for three degree of freedom (that is three independent variables) are calculated as

   \begin{eqnarray}\label{eq233}
 \chi^2=\frac{1}{3} \left[\frac{(a_m-a_T)^2}{\Delta a_m^2+ \Delta a_T^2} + \frac{(b_m-b_T)^2}{\Delta b_m^2+ \Delta b_T^2} + \frac{(c_m-c_T)^2}{\Delta c_m^2+ \Delta c_T^2} \right],
\end{eqnarray}
 where the $\Delta$ values indicate the measured regression errors. We found $\chi^2\gg1$ for all models: a fact that  proves that all GCMs fail to simultaneously reproduce the 20-year, 60-year and the upward trend observed in the temperature with a probability higher than 99.9\%. This $\chi^2$ measure based on the multidecadal patterns is quite important because climate changes on a multidecadal scale are usually properly referred to as \emph{climate changes}, and a climate model should at least get these temperature variations right to have any practical economical medium-range planning utility such as street construction planning, agricultural and industrial location planning, prioritization of scientific energy production research versus large scale applications of current very expensive green energy technologies, etc.

 It is also possible to include in the discussion the two detected decadal cycles as
 \begin{equation}\label{eq77}
    g(t)=C_3\cos\left[\frac{2\pi(t-T_3)}{10.44}\right]  +C_4\cos\left[\frac{2\pi(t-T_4)}{9.07}\right].
\end{equation}
A detailed discussion about the  choice of the two above periods and their physical meaning is better addressed  in Section 4.
 Fitting the temperature for the period 1850-2011 gives: $C_3=0.03\pm0.01~^oC$, $T_3=2002.7\pm0.5~AD$, $C_4=0.05\pm0.01~^oC$, $T_4=1997.7\pm0.3~AD$. It is possible to test how well the IPCC GCMs reconstruct these two decadal cycles by fitting their simulations with the following equation

\begin{equation}\label{eq266}
    n(t) = m(t) + s*0.03\cos\left[\frac{2\pi (t-2002.7)}{10.44}\right]+ l*0.05\cos\left[\frac{2\pi (t-1997.7)}{9.07}\right] ~,
\end{equation}
where $s$ and $l$ are  regression coefficients. Values of $s$ and $l$ statistically compatible with the number 1 indicate that the model well reproduces the two observed decadal temperature cycles, respectively.  On the contrary, values of $s$ and $l$ statistically incompatible with 1  indicate that the model does not reproduce the observed temperature cycles. The results referring the average model run, as defined above, are reported in Table 2, where it is evident that the GCMs fail to reproduce these two decadal cycles as well. The average values among the 26 models is:  $s=0.06\pm0.40$ and $l=0.34\pm0.37$, which are statistically different from 1. In many cases the regression coefficients are even negative. The table also includes the reduced $\chi^2$ (chi square) values for five  degree of freedom by extending Eq. \ref{eq233} to include the other two decadal cycles. Again, we found $\chi^2\gg1$ for all models.

Finally, we can  estimate  how well the astronomical model made of the sum of the four harmonics plus the quadratic trend (that is: $f(t)+g(t)+p(t)$) reconstructs the 1850-2011 temperature record relative to the GCM simulations. For this purpose we
evaluate the root mean square (RMS) residual values between the 4-year average smooth curves of each GCM average simulation and the 4-year average smooth of the temperature curve, and we do the same between the astronomical model and the 4-year average smooth temperature curve. We use a 4-year average smooth because the model is not supposed to reconstruct the fast sub-decadal fluctuations.  The RMS residual values are reported in Table 2. The RMS residual value relative to the harmonic model is 0.051 $^oC$, while for the GCMs we get RMS residual values from 2 to 5 times larger. This result further indicates that the geometrical model is significantly more accurate than the GCMs in reconstructing the global surface temperature from 1850 to 2011.

The above finding reinforces the conclusion of Scafetta (2010b) that the IPCC (2007) GCMs do not reproduce the observed major decadal and multidecadal dynamical patterns observed in the global surface temperature record. This conclusion does not  change if the single GCM runs are studied.

\section{Reconstruction of the global surface temperature oscillations: 1880-2011}

A regression model may always produce results in a reasonable agreement within the same time interval used for its calibration. Thus, showing that an empirical  model can reconstruct the same data used for determining its free regression parameters would be not surprising, in general. However, if the same model is shown to be capable of forecasting the patterns of the data outside the temporal interval used for its statistical calibration, then the model likely has a physical meaning. In fact, in the later case the regression model would be using constructors that are not simply independent generic mathematical functions, but are functions that capture the dynamics of the system under study. Only a mathematical model that is shown to be able to both reconstruct and forecast (or predict) the observations is physically relevant according the scientific method.

The climate reconstruction efficiency of an empirical climate model based on a set of astronomical cycles with the periods herein analyzed has been tested and verified in \cite{scafettanew}, \cite{Loehle} and \cite{scafett2011b}. Herein, we simply summarize some results for the benefit of the reader and for introducing the following section.

In figures 10 and 11 in \cite{scafettanew} it is shown that the 20-year and 60-year oscillations of the speed of the Sun relative to the barycenter of the solar system are in a very good phase synchronization with the correspondent 20 and 60-year climate oscillations. Moreover, detailed spectra analysis has revealed that the climate system shares numerous other frequencies with the astronomical record.

In figures 3 and 5 in \cite{Loehle} it is shown that an harmonic model based on 20-year and 60-year cycles and free phases calibrated on the global surface temperature data for the period 1850-1950 is able to properly reconstruct the 20-year and 60-year  modulation of the temperature observed since 1950. This includes a small peak around 1960, the cooling from 1940 to 1970, the warming from 1970 to 2000 and a slight stable/cooling trending since 2000. It was also found a quasi linear residual with a warming trending of about $0.66 \pm 0.16 ~^oC/century$ that was interpreted as due to a net anthropogenic warming trending.

In \cite{scafett2011b}, it was found that the historical mid-latitude aurora record, mostly from central and southern Europe, presents the same major decadal and multidecadal oscillations of the astronomical records and of the global surface temperature  herein studied. It has been shown that  a harmonic model with aurora/astronomical cycles with periods of  9.1, 10.5, 20, 30 and 60 years calibrated during the period 1850-1950 is able to carefully reconstruct the decadal and multidecadal oscillations of the temperature record since 1950. Moreover, the same harmonic model calibrated during the period 1950-2010 is able to carefully reconstruct the decadal and multidecadal oscillations of the temperature record from 1850 to 1950. The argument about the 1850-1950-fit versus 1950-2010-fit is crucial for showing the forecasting capability of the proposed harmonic model. This property is what distinguishes a mere curve fitting exercise from a valid empirical dynamical model of a physical system. This is a major requirement of the scientific method \cite{scafett2011b}.
A preliminary physical model based on a forcing of the cloud system has been proposed to explain the synchrony between the climate system and the astronomical oscillations.

The above results have supported the thesis that climate is forced by astronomical oscillations and can be partially reconstructed and forecasted by using the same cycles, but for an efficient forecast there is the need of additional information. This is done in the next section.

\section{Corrected anthropogenic projected warming trending and forecast of the global surface temperature: period 2000-2100}

Even assuming that the detected decadal and multidecadal cycles will continue in the future, to properly forecast climate variation for the next decades, additional information  is necessary: 1) the amplitudes and the phases of possible multisecular and millennial cycles; 2)  the net anthropogenic contribution to the climate warming according to realistic emission scenarios.

The first issue is left to another paper because it requires a detailed study of the paleoclimatic temperature proxy reconstructions which are relatively different from each other. These cycles are those responsible for the cooling periods during the Maunder and Dalton solar minima as well as for the Medieval Warm Period and the Little Ice Age. So, we leave out these cycles here. Considering that we may be at the very top of these longer cycles, ignoring their contribution may be reasonable only if our forecast is limited  to the first decades of the 21$^{st}$ century. However, a rough preliminary estimate would suggests that these longer cycles may contribute globally to an additional cooling of about 0.1-0.2 $^oC$ by 2100 because the millenarian cycle presents an approximate min-max amplitude of about 0.5-0.7 $^oC$ \citep{Ljungqvist} and the top of these longer cycles would occur somewhere during the 21$^{st}$ century  \citep{Humlum,Liu}. Secular and millennial longer natural cycles could have contributed about 0.2-0.3 $^oC$ warming from 1850 to 2010 (Scafetta and West, 2007; Eichler et al., 2009: Scafetta, 2009, 2010a).

The second issue is herein explicitly addressed by using an appropriate argument that adopts the same GHG emission scenarios utilized by the IPCC, but correct their climatic effect.
In fact, the combination of the 20-year and 60-year cycles, as evaluated in Eq. \ref{eq1}, should have contributed for about $0.3~^oC$ of the 0.5 $^oC$ warming observed from 1970 to 2000.
 During this period the IPCC (2007) have claimed, by using the GCMs studied herein, that the natural  forcing (solar plus volcano) would have caused a cooling up to 0.1-0.2 $^oC$ (see figure 9.5b in the IPCC report, which is herein reproduced with added comments in Figure S3A in the the Section 4 in the Supplement file). As it is evident in the IPCC figure 9.5a (also shown in the Supplement file), the IPCC GCM results imply that from 1970 to 2000 the net anthropogenic forcing contributed a net warming of the observed 0.5 $^oC$ plus, at most, another 0.2 $^oC$, which had to offset the alleged natural volcano cooling of up to -0.2 $^oC$. A 0.7 $^oC$ anthropogenic warming trend in this 30-year period corresponds to an average anthropogenic warming rate of about $2.3 ^oC/century$ since 1970. This  value is a realistic estimate of the average GCM performance because the average GCM projected anthropogenic net warming rate is $2.3\pm0.6~ ^oC/century$  from 2000 to 2050 according to several GHG emission scenarios (see figure SPM.5 in the IPCC report, which is herein reproduced with added comments in Figure S4B in the Supplement file).

  On the contrary, if about $0.3~^oC$ of the warming observed from 1970 to 2000 has been naturally induced by the 60-year natural modulation during its warming phase, at least 43-50\% of the alleged 0.6-0.7 $^oC$ anthropogenic warming has been naturally induced, and the 2.3 $^oC/century$ net anthropogenic trending should be reduced at least to 1.3 $^oC/century$.

  However, the GCM alleged 0.1-0.2 $^oC$ cooling from 1970 to 2000 induced by volcano activity may be a gross overestimation of the reality.  In fact, as revealed in Figure 2,  the GCM climate simulation presents a strong volcano signature peak at 80-year time lag that is totally absent in the temperature record, even after filtering. This would imply that the volcano signature should be  quite smaller and shorter than what the GCMs estimate, as  empirical studies have shown \citep{Lockwood,Thompson}. Section  5 of the Supplement file  shows that the GISS ModelE appears to greatly overestimate the long-time signature associated to volcano activity against the same signature as estimated by empirical studies.

  Moreover, the observed 0.5 $^oC$ warming from  1970 to 2000, which the IPCC models associate to anthropogenic GHG plus aerosol emissions and to other anthropogenic effects, may also be partially due to poorly corrected urban heat island (UHI) and land use changes (LUC) effects, as argued in detailed statistical studies \citep{McKitrick1,McKitrick2}. As extensively discussed in those papers, it may  be reasonable that the $\sim0.5 ~^oC$ warming reported since 1950-1970 in the available temperature records has been overestimated up to 0.1-0.2 $^oC$ because of poorly corrected  UHI and LUC effects. Indeed, the land warming since 1980 has been almost twice the ocean warming, which may be not fully explained by the different heat capacity between land and ocean. Moreover, during the last decades the agencies that provide the global surface temperature records have changed several times the methodologies adopted to attempt to correct UHI and LUC spurious warming effects and, over time, have produced quite different records \citep{Daleo}. Curiously, the earlier  reconstructions show a smaller global warming and a more evident 60-year cyclical modulation from 1940 to 2000 than the most recent ones.

  Finally, there may be an additional natural warming due to multisecular and millennial cycles as   explained in the Introduction. In fact, the solar activity increased during the last four centuries \citep{Scafetta3}, and the observed global surface warming during the 20th century is very likely also part of a natural and persistent recovery from the Little Ice Age of AD 1300-1900  \citep{Scafetta2007,Scafetta3,Loehle,Soon2009,Soon2011,Kirkby,Hoyt,Courtillot44,Thejll,Weihong,Eichler,Humlum,Liu}: see also section 7 in the Supplement file.

Thus, the above estimated 1.30 $^oC/century$ anthropogenic warming trending is likely an upper limit estimate. As a lower limit we can reasonably assume the $0.66\pm0.16~^oC/century$, as estimated in \cite{Loehle}, which would be compatible with the claim that only 0.2 $^oC$ warming (instead of 0.7 $^oC$) of the observed 0.5 $^oC$ warming since 1970 could be anthropogenically  induced. This result would be consistent with the fact that according empirical studies \citep{Lockwood,Thompson}  the cooling long-range effects of the volcano eruptions almost vanished in 2000 (see Section 5 in the Supplement file) and that the secular natural trend could still be increasing. So, from 2000 to 2050 we claim that the same IPCC (2007) anthropogenic emission projections could only induce a warming trend approximately described by the curve

\begin{equation}\label{}
    q(t)=(0.009\pm 0.004) (t-2000).
\end{equation}

There are also  two major quasi decadal oscillations with periods of about 9.1 yr and 10-10.5 yr: see Figure 3.
The 9.1-year cycle may be due to a Soli/Lunar tidal cycle \citep{scafettanew,scafett2011b} . In fact, the lunar apsidal line rotation period is 8.85 years while the Soli/Lunar nodal cycle period is 18.6 years. Note that there are two nodes and the configuration Sun-Moon-Earth and Sun-Earth-Moon are equivalent for the tides: thus, the resulting tidal cycles should have a period of about 18.6/2=9.3 yr. The two cycles at 8.85-year and 9.3-year should beat, and produce a fast cycle with an average period of $2/(1/8.85+1/9.93)=9.07$ yr that could be modulated by a slow cycle with period  of   $2/(1/8.85-1/9.93)=182.9$ yr. There may also be an additional influence of the half Saros eclipse cycle that is about 9 years and 5.5 days. In conclusion, the quasi 9.1-year cycle appears to be related to a Soli/Lunar tidal cycle dynamics. The 10-10.5-year cycle has been interpreted as related to an average cycle between the $0.5/(1/11.862-1/29.457)=9.93$ yr Jupiter/Saturn half-synodic tidal cycle and the 11-year solar cycle (we would have a beat cycle with period of $2/(1/9.93+1/11)=10.44$ yr). Moreover, a quasi 9.91-year and 10.52-year cycles have been found in the natural gravitational resonances of the solar system \citep{Bucha,Grandpiere,scafett2011b}.

It is possible to include these two cycles in the harmonic model using the additional harmonic function Eq. (\ref{eq77})
and our final model based on 4-frequency harmonics plus two independent trending functions is made as
\begin{equation}\label{hm}
  h(t)= f(t)+ g(t)+ \left\{ \begin{array}{ll}
 p(t)&\mbox{ if $1850<t<2000$} \\
  p(2000) + q(t) &\mbox{ if $2000<t<2100$}
       \end{array} \right.
\end{equation}

To test the forecasting capability of the $g(t)$ harmonics, the $f(t)+g(t)+p(t)$ model is calibrated in two complementary periods. Note that $g(t)$ is sufficiently orthogonal to $f(t)+p(t)$, so we keep $f(t)+p(t)$ unchanged for not adding too many free regression parameters.   Fitting the period 1850-1950 gives: $C_3=0.03\pm0.01~^oC$, $T_3=2003\pm0.5~AD$, $C_4=0.05\pm0.01~^oC$, $T_4=1997.5\pm0.3~AD$. Fitting the period 1950-2011 gives: $C_3=0.04\pm0.01~^oC$, $T_3=2002.1\pm0.5~AD$, $C_4=0.05\pm0.01~^oC$, $T_4=1998.1\pm0.3~AD$. Fitting the period 1850-2011 gives: $C_3=0.03\pm0.01~^oC$, $T_3=2002.7\pm0.5~AD$, $C_4=0.05\pm0.01~^oC$, $T_4=1997.7\pm0.3~AD$. If the decadal period 10.44 yr is substituted with a 10 yr period for 1850-2011, we get: $C_3=0.02\pm0.01~^oC$, $T_3=2000.4\pm0.5~AD$, $C_4=0.04\pm0.01~^oC$, $T_4=1997.7\pm0.3~AD$.

We observe that all correspondent amplitudes and phases coincide within the error of measure, which implies that the model has forecasting capability. Moreover, the phase related to the 9.1-year cycle presents a maximum around 1997-1998. We observe that this period is in good phase with the Soli/Lunar nodal dates at the equinoxes, when the Soli/Lunar spring tidal maxima are located in proximity of the equator,  and the  extremes in the tidal variance  occurs \citep{Sidorenkov}. In fact, each year there are usually two solar eclipses and two lunar eclipses, but the month changes every year and the cycle repeats every about 9 years with the moon occupying the opposite node. Thus, eclipses occur, within a two week interval, close to the equinoxes (around March 20/21 and September 22/23) every almost 9 years. Section 6 in the Supplement file reports the dates of the solar and lunar eclipses occurred from 1988 to 2010 and compares these dates with the detected 9-year temperature cycle. Two lunar eclipses occurred on 24/Mar/1997 and 16/Sep/1997, the latter eclipse also occurred at the lunar perigee (that is, when the Moon is in its closest position to the Earth) so that  the line of the lunar apsides too was oriented along the Earth-Sun direction (so that the two cycles could interfere constructively).  Two solar eclipses took place almost 9-years later at almost the same dates, 22/Sep/2006 (at the lunar apogee) and 19/Mar/2007 (at the lunar perigee). This date matching   suggests that the 9.1-year cycle is likely related to a Soli/Lunar tidal cycle. Indeed, this cycle is quite visible in the ocean oscillations \citep{scafettanew,scafett2011b} and ocean indexes such as the Atlantic Multidecadal Oscillation (AMO) and the Pacific Decadal Oscillation (PDO).

The timing of the 10-10.5-year cycle maximum (2000-2003), corresponds relatively well with the total solar irradiance maximum in 2002 \citep{scafettwill} and the Jupiter/Saturn conjunction around 2000.5 (so that the two cycles could interfere constructively). This suggests that this decadal cycle has a solar/astronomical origin.

The above information is combined in Figures 5A and 5B that depict: the monthly sampled global surface temperature since  1850; a 4-year moving average estimates of the same; the proposed model given in Eq.  \ref{hm} with two and four cycles, respectively.  Finally, for comparison, we plot  the IPCC projected  warming using the average GCM projection estimates, which is given by  a linear trending warming of $2.3\pm0.6~^oC/century$ from 2000 to 2050 while since 2050 the projections spread a little bit more according to alternative emission scenarios (see figure S4B in Section 4 in the Supplement file). The two figures are complementary by highlighting both a low resolution forecast that extends to 2100, which can be more directly compared with the IPCC projections, and a higher resolution forecast for the next decades that may be more important for an immediate economical planning, as explained above.

\begin{figure}
\begin{center}
\includegraphics[angle=0,width=30pc]{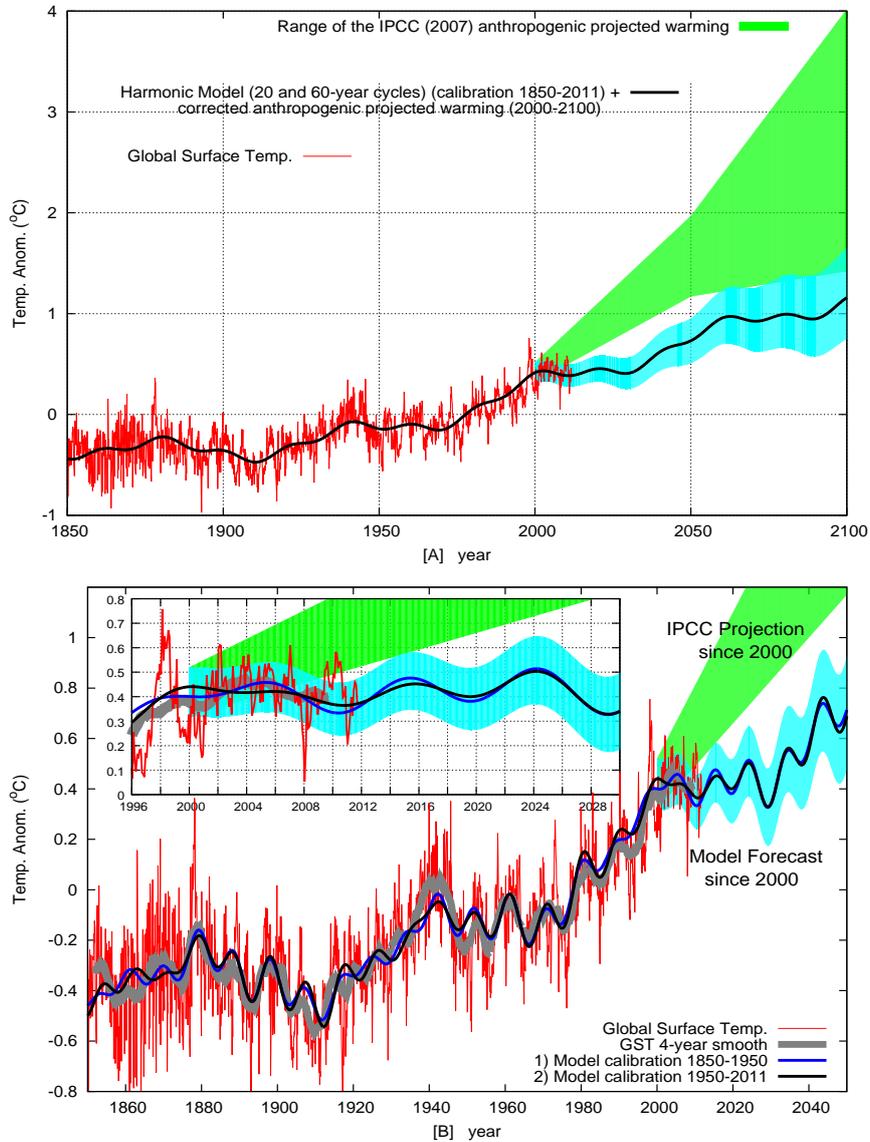}
\end{center}
 \caption{[A] The monthly sampled global surface temperature from 1950 to 2050 (red); the proposed empirical model (Eq.  \ref{hm}) made of the discussed 2 cycles (20 and 60 yr) plus the quadratic trend until 2000 that is substituted with the corrected anthropogenic net projected warming as explained in the text (black); the IPCC 2007 projections (green).  [B] The monthly sampled global surface temperature from 1950 to 2050 (red); a 4-year moving average estimates of the same (smooth wide gray curve); the proposed empirical model (Eq.  \ref{hm}) made of the discussed 4 cycles (9.07, 10.44, 20 and 60 yr) plus the quadratic trend until 2000 that is substituted with the anthropogenic net estimated contribution given by a linear trend with a rate within the interval 0.5-1.3 $^oC/century$ as discussed in the text (black and blue small curves);  finally, by comparison  the IPCC projected warming using the average GCM projection with a trend of $2.3\pm0.6~^oC/century$ from 2000 to 2050. Note than the two harmonic model curves  use the two decadal harmonics at 9.07-year and 10.44-year periods calibrated on the temperature data during two complementary time periods, 1850-1950 and 1950-2011 respectively. As evident in the figure, the decadal oscillations reconstructed by the two alternative models are very well synchronized between them and with the oscillations revealed in the grey 4-year smooth temperature gray curve. This result suggests that the astronomical harmonic model has forecast capability. The insert figure is reproduced in a full page figure in the supplement file.}
\end{figure}

Figure 5 clearly shows the good performance of the proposed model (Eq. \ref{hm}) in reconstructing the dacadal and multidecadal oscillations of the global surface temperature since 1850.
The model has forecasting capability also at the decadal scale because the two curves  calibrated using the independent  periods 1850-1950 and 1950-2011 are synchronous to each other also at the decadal scale and are synchronous with the temperature modulation revealed by the 4-year smooth curve: the statistical divergence between the harmonic model reconstruction  and the data have a standard deviation of $\sigma=0.15~^oC$, which is due to the large and fast ENSO related oscillations, while the divergence with the grey 4-year smooth curve of the temperature has a standard deviation of $\sigma=0.05~^oC$, as Table 2 reports.

Figure 5 shows that the IPCC warming projection since 2000 (at a rate of $2.3+0.6~^oC/century$ plus a vertical error of $\pm0.1 ~^oC$ ) does not  agree with the observed temperature pattern since about 2005-2006.
 On the contrary, the empirical model we propose, Eq.  \ref{hm}, appears to reasonably forecast the observed trending of the global surface temperature since 2000, which appears to have been almost steady: the error bars are calculated by taking into account both the statistical error of the model ($\sim \pm0.1~^oC$) (because, at the moment, the harmonic model includes only the decadal and multidecadal scales and, evidently, it is  not supposed to reconstruct the fast ENSO related oscillations) plus the projected anthropogenic net warming with a linear rate within the interval 0.5-1.3 $^oC/century$, as discussed above. According our model, by 2050 the climate may warm by about 0.1-0.5 $^oC$, which is significant less than the average $1.2 \pm 0.4 ~^oC$ projected by the IPCC. If multisecular natural cycles (which according to several authors have significantly contributed to the observed 1700-2000 warming  and very likely will contribute to a cooling since the 21$^{st}$ century)   are ignored, the temperature may warm by about 0.3-1.2 $^oC$ by 2100 contrary to the 1.0-3.6 $^oC$ warming projected by the IPCC (2007) according to its various emission scenarios.

The divergence of the temperature data from the IPCC projections and their persistent convergence with the astronomical harmonic model  can be calculated by evaluating a time continuous discrepancy $\chi^2(t)$ (\emph{chi-squared}) function defined as

\begin{equation}\label{rrr}
    \chi^2 (t)=\frac{(Tem(t)-Mod(t))^2}{(\Delta Mod(t))^2},
\end{equation}
where $Tem(t)$ is the 4-year smooth average temperature curve depicted in the figure, which highlights the decadal oscillation,  $Mod(t)$ is used first for indicating  the IPCC GCM average projection curve and second for indicating the harmonic model average forecast curve as depicted in the figure, and $\Delta Mod(t)$ is used to indicate the time dependent uncertainty first of the IPCC projection and second of the harmonic model, respectively, which are depicted in the two shadow regions in Figure 5. In the above equation the implicit error associated to the 4-year smooth average temperature curve is considered negligible (it has an order of magnitude of 0.01 $^oC$) compared to the uncertainty of the models $\Delta Mod(t)$, which has an order of magnitude of 0.1 $^oC$ and above, so we can ignore it in the denominator of Eq. \ref{rrr}.   Values of $\chi^2 (t)<1$ indicate a sufficient \emph{agreement} between the data and the model at the particular time $t$, while values of $\chi^2 (t)>1$ indicate \emph{disagreement}. Figure 6 depicts Eq. \ref{rrr} and clearly shows that the astronomical harmonic model forecast is quite accurate as the time progress since 2000. Indeed, the performance of our geometrical model   is always superior than the IPCC projections. The IPCC (2007) projections significantly diverge from the data since 2004-2006.

\begin{figure}
\begin{center}
\includegraphics[angle=0,width=30pc]{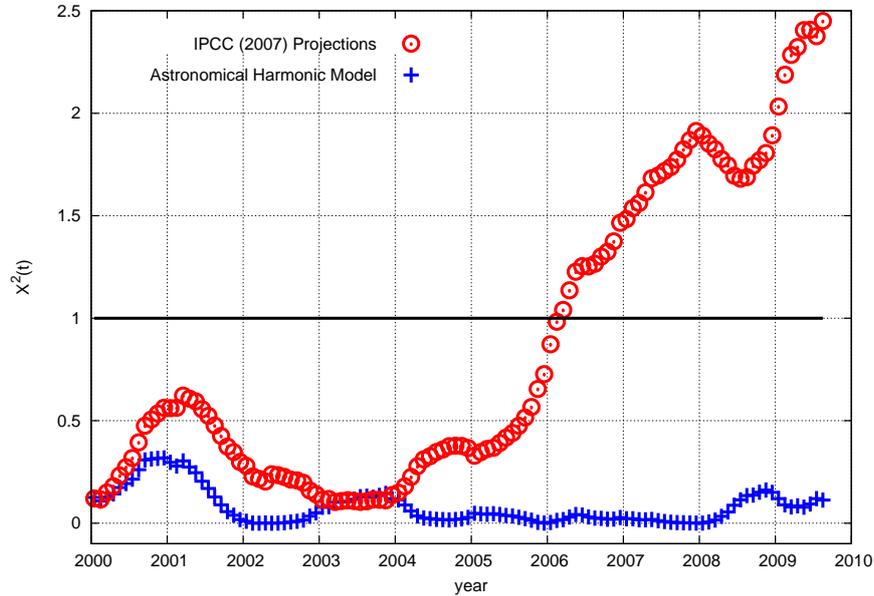}
\end{center}
 \caption{ The curves are produced with  Eq. \ref{rrr} and give a time dependent estimate of how well the astronomical harmonic model (crosses) and the IPCC projections (circles) forecast the temperature data (4-year smooth average temperature gray curve in Figure 5) since 2000. The IPCC projections significantly diverge from the temperature data since 2004-2006. The astronomical harmonic model is shown to forecast the data quite well and it is quite stable in time. The variable $\chi^2$ needs to be less than 1 for statistical compatibility.}
\end{figure}

\section{Discussion and Conclusion}

The scientific method requires that a physical model fulfils two conditions: it has to reconstruct \emph{and} predict (or forecast) physical observations.
Herein, we have found that the GCMs used by the IPCC (2007) seriously fail to properly reconstruct even the large multidecadal oscillations found in the global surface temperature which have a climatic meaning.
  Consequently, the IPCC projections for the $21^{st}$ century cannot be trusted.  On the contrary, the astronomical empirical harmonic model proposed in   \cite{scafettanew,scafett2011b} has been shown to be capable of reconstructing and, more importantly, forecasting the decadal and multidecadal oscillations found in the global surface temperature with a sufficiently good accuracy.  Figures 5 and 6 shows that in 1950 it could have been possible to accurately forecast the decadal and multidecadal oscillations observed in the climate since 1950, which includes a steady/cooling trend from 2000 to 2011. Four major cycles have been detected and used herein with period of 9.1 yr (which appears to be linked to a Soli/Lunar tidal cycle), and of 10-10.5, 20-21 and 60-61 yr (which appears to be  in phase with the gravitational cycles of Jupiter and Saturn that can also modulate the solar cycles at the equivalent time-scales). However, other astronomical cycles may be involved in the process.

  This result argues in favor of a celestial origin of the climate oscillations   and whose mechanisms were not included in the climate models adopted by the IPCC in 2007. The harmonic interpretation of climate change also appears more reasonable than recent attempts of reproducing with GCMs some limited climate pattern such as the observed slight cooling from 1998 to 2008 by claiming that it is a red-noise-like internal fluctuation of the climatic system \citep{Meehl} or by carefully playing with the very large uncertainty in the climate sensitivity to $CO_2$ changes and in the aerosol forcing \citep{Kaufmann}. In fact, a quasi 60-year cycle in the climate system has been observed for centuries and millennia in several independent records, as explained in the Introduction.

  By not properly reconstructing the 20-year and 60-year natural cycles we found that the IPCC GCMs have seriously overestimated also the   magnitude of the anthropogenic contribution to the recent global warming.
  Indeed, other independent studies have found serious incompatibilities between the IPCC climate models and the actual observations and reached the same conclusion. For example, \cite{Douglass}   showed that there is a large discrepancy between observed tropospheric temperature trends and the IPCC climate model predictions from Jan 1979 to Dec 2004: GCM ensemble mean simulations show that the increased $CO_2$ concentration should have produced an increase in the tropical warming trend with altitude, but balloon and satellite observations do not show any increase \citep{Singer}.   \cite{Spencer} have showed that there is a large discrepancy between the satellite observations and the behavior of the IPCC climate models on how the Earth loses energy as the surface temperature changes. Both studies imply that the modeled climate sensitivity to $CO_2$ is largely overestimated by the IPCC models.
  Our findings would be consistent with the above results too and would imply a climate sensitivity to $CO_2$ doubling much lower than the IPCC's proposal of 1.5-4.5 $^oC$. \cite{Lindzen} has argued for a climate sensitivity to a $CO_2$ doubling of 0.5 $^oC$ - 1.3 $^oC$ by using variations in Earth's radiant energy balance as measured by satellites.
We claim that the reason of the discrepancy between the model outcomes and the data is  due  to the fact  that the current GCMs are missing major astronomical forcings related to the harmonies of the solar system and the physical/climatic mechanisms related to them \citep{scafett2011b}.

Probably several solar and terrestrial mechanisms are involved in the process \citep{Scafetta3,scafettanew,scafett2011b}. It is reasonable that with their gravitational and magnetic fields, the planets can directly or indirectly modulate the solar activity, the heliosphere, the solar wind and, ultimately, the terrestrial magnetosphere and ionosphere.
In fact, planetary tides, as well as solar motion induced by planetary gravity may increase solar nuclear fusion rate \citep{Grandpiere,Wolff}. Moreover,  \cite{Charvatova}, \cite{Komitov}, \cite{Mazzarella} and \cite{scafett2011b} showed that the historical multi-secular aurora record and some cosmogenic beryllium records presents a large quasi 60-year cycle which would suggest that the astronomical cycles regulated by Jupiter and Saturn are the primary indirect cause of the oscillations in the terrestrial ionosphere. \cite{Ogurtsov} have found that several multi secular solar reconstructions do present a quasi 60-year cycle together with longer cycles. \cite{Loehle} have argued that a quasi 60-year cycle may be present in the total solar irradiance (TSI) since 1850, although the exact reconstruction of TSI is not currently possible.
Indeed, TSI direct satellite measurements since 1978 have produced alternative composites such as the ACRIM \citep{Willson}, which may present a pattern that would be compatible with a 60-year cycle. In fact, the ACRIM TSI satellite composite presents an increase from 1980 to  2002 and a decrease afterward. On the contrary, the PMOD TSI composite adopted by the IPCC \cite{Frohlich} does not present any patter resembling a 60-year modulations but a slightly decrease since 1980. However, the way how the PMOD science team has adjusted the TSI satellite  records to obtain its composite may be erroneous \citep{scafettwill,scafettwill22}.

 Indeed, \cite{scafett2011b} found that several mid-latitude aurora cycles (quasi 9.1, 10-10.5, 20-21 and 60-62 yr cycles) correspond to the climate cycles herein detected.
We believe that the oscillations found in the historical mid-latitude aurora record are quite important because  reveal the existence of   equivalent oscillations in the electric properties  of the atmosphere, which  can regulate the cloud system \citep{Svensmark98,Carslaw,Svensmark,Tinsley,Kirkby,Enghoff,Kirkby11}. In addition, the variations in solar activity also modulate the incoming cosmic ray flux that may lead to a cloud modulation.  The letter too would modulate the terrestrial albedo with the same frequencies found in the solar system.  As shown in Scafetta (2011b) just a 1-2\% modulation of the albedo would be sufficient to reproduce the climatic signal at the surface, which is an amplitude compatible with the observations. Oscillations in the albedo would cause correspondent oscillations in the climate mostly through warming/cooling cycles induced in the ocean surface. For example, a 60-year modulation has been observed in the frequency of  major hurricanes on the Atlantic ocean that has been associated to a 60-year cycle in the strength of the Atlantic Thermohaline Circulation (THC), which would also imply a similar oscillation in the Great Ocean Conveyor Belt \citep{Gray}. Moreover, herein we have found further evidences that the 9.1-year cycle is linked to the Soli/Lunar tidal dynamics.
Ultimately, the climate amplifies the effect of harmonic forcing through several internal feedback mechanisms, which ultimately tend to synchronize all climate oscillations with the solar-lunar-planetary astronomical oscillations through collective synchronization mechanisms \citep{Pikovsky,Strogatz,scafettanew}.

For the above reasons, it is very unlikely that the observed  climatic oscillations are due only to an internal variability of the climate system that evolves independently of astronomical forcings,  as proposed by some authors \citep{Latif,Meehl}. Indeed, the GCMs do not really reconstruct the actual observed oscillations at all temporal scales, nor they have ever been able to properly forecast them. It is evident that simply showing that a  model is able to produce some kind of red-noise-like variability  (as shown in the numerous GCM simulations depicted in the figures in the Supplement file) is not enough to claim that the model has really modeled the observed dynamics of the climate.

For the imminent future, the global climate may remain approximately steady until 2030-2040, as it has been observed from the 1940s to the 1970s because the 60-year climate cycle has entered into its cooling phase around 2000-2003, and this cooling will oppose the adverse effects of a realistic anthropogenic global warming, as shown in Figure 5. By using the same IPCC projected anthropogenic emissions our partial empirical harmonic model forecast forecast a global warming by about 0.3–1.2 $^oC$ by 2100, contrary to the IPCC 1.0–3.6 $^oC$ projected warming.  The climate may also further cool if additional natural secular and millennial cycles enter into their cooling phases. In fact, the current warm period may be part of a quasi millennial natural cycle, which is currently at its top as it was during the roman and medieval times, as can be deduced from climate records \citep{Schulz,Ljungqvist} and solar records covering the last millennia \citep{Bard,Ogurtsov}. Preliminary attempts to address this issue have been made by numerous authors as discussed in the Introduction such as, for example, by  \cite{Humlum}, while  a more detailed discussion  is left to another  paper.

\begin{table}
  \centering
\begin{tabular}{|c|l|c|c|c|c|c|}
  \hline
\#	&	model	&	a		(60-year)	&	b		(20-year)	&	c		(trend)	&	d (bias)			 &	 $\chi^2$	(abc) \\ \hline
	&	temp	&	1.03	$\pm$	0.05	&	0.99	$\pm$	0.12	&	1.01	$\pm$	0.02	&	0.00	$\pm$	 0.01	&	0.21	\\ \hline
1	&	GISS ModelE	&	0.25	$\pm$	0.03	&	0.90	$\pm$	0.08	&	0.80	$\pm$	0.01	&	0.08	 $\pm$	0.01	&	89	\\ \hline
2	&	BCC CM1	&	0.63	$\pm$	0.03	&	0.69	$\pm$	0.09	&	0.54	$\pm$	0.02	&	0.08	$\pm$	 0.01	&	109	\\ \hline
3	&	BCCR BCM2.0	&	0.29	$\pm$	0.05	&	0.06	$\pm$	0.11	&	0.40	$\pm$	0.02	&	0.08	 $\pm$	0.01	&	202	\\ \hline
4	&	CGCM3.1 (T47)	&	0.35	$\pm$	0.03	&	-0.28	$\pm$	0.07	&	2.02	$\pm$	0.01	&	0.40	 $\pm$	0.01	&	753	\\ \hline
5	&	CGCM3.1 (T63)	&	0.11	$\pm$	0.05	&	0.05	$\pm$	0.11	&	2.07	$\pm$	0.02	&	0.40	 $\pm$	0.01	&	536	\\ \hline
6	&	CNRM CM3	&	-0.01	$\pm$	0.07	&	-0.27	$\pm$	0.18	&	2.02	$\pm$	0.03	&	0.39	 $\pm$	0.01	&	322	\\ \hline
7	&	CSIRO MK3.0	&	0.30	$\pm$	0.04	&	-0.12	$\pm$	0.11	&	0.48	$\pm$	0.02	&	0.08	 $\pm$	0.01	&	176	\\ \hline
8	&	CSIRO MK3.5	&	-0.19	$\pm$	0.04	&	-0.19	$\pm$	0.10	&	1.38	$\pm$	0.02	&	0.25	 $\pm$	0.01	&	197	\\ \hline
9	&	GFDL CM2.0 	&	0.44	$\pm$	0.05	&	0.90	$\pm$	0.12	&	1.12	$\pm$	0.02	&	0.21	 $\pm$	0.01	&	28	\\ \hline
10	&	GFDL CM2.1 	&	0.37	$\pm$	0.07	&	0.75	$\pm$	0.17	&	1.37	$\pm$	0.03	&	0.26	 $\pm$	0.01	&	53	\\ \hline
11	&	GISS AOM	&	0.22	$\pm$	0.03	&	-0.14	$\pm$	0.06	&	1.10	$\pm$	0.01	&	0.22	 $\pm$	0.01	&	93	\\ \hline
12	&	GISS EH	&	0.48	$\pm$	0.04	&	0.96	$\pm$	0.11	&	0.80	$\pm$	0.02	&	0.14	$\pm$	 0.01	&	43	\\ \hline
13	&	GISS ER	&	0.47	$\pm$	0.04	&	0.80	$\pm$	0.08	&	0.90	$\pm$	0.02	&	0.11	$\pm$	 0.01	&	31	\\ \hline
14	&	FGOALS g1.0	&	0.10	$\pm$	0.09	&	-0.15	$\pm$	0.21	&	0.28	$\pm$	0.03	&	0.06	 $\pm$	0.01	&	171	\\ \hline
15	&	INVG ECHAM4	&	-0.12	$\pm$	0.05	&	0.37	$\pm$	0.12	&	1.34	$\pm$	0.02	&	0.24	 $\pm$	0.01	&	138	\\ \hline
16	&	INM CM3.0	&	0.30	$\pm$	0.07	&	0.47	$\pm$	0.18	&	1.34	$\pm$	0.03	&	0.24	 $\pm$	0.01	&	54	\\ \hline
17	&	IPSL CM4	&	0.13	$\pm$	0.06	&	0.05	$\pm$	0.14	&	1.37	$\pm$	0.02	&	0.26	 $\pm$	0.01	&	107	\\ \hline
18	&	MIROC3.2 Hires	&	0.35	$\pm$	0.05	&	0.92	$\pm$	0.12	&	1.43	$\pm$	0.02	&	0.19	 $\pm$	0.01	&	104	\\ \hline
19	&	MIROC3.2 Medres	&	0.34	$\pm$	0.03	&	0.76	$\pm$	0.09	&	0.72	$\pm$	0.01	&	0.14	 $\pm$	0.01	&	104	\\ \hline
20	&	ECHO G	&	0.58	$\pm$	0.04	&	0.16	$\pm$	0.10	&	0.98	$\pm$	0.02	&	0.18	$\pm$	 0.01	&	26	\\ \hline
21	&	ECHAM5/MPI-OM	&	0.19	$\pm$	0.04	&	0.31	$\pm$	0.09	&	0.70	$\pm$	0.02	&	-0.02	 $\pm$	0.01	&	104	\\ \hline
22	&	MRI CGCM 2.3.2	&	0.31	$\pm$	0.03	&	0.03	$\pm$	0.07	&	1.36	$\pm$	0.01	&	0.27	 $\pm$	0.01	&	149	\\ \hline
23	&	CCSM3.0	&	0.34	$\pm$	0.04	&	0.43	$\pm$	0.10	&	1.29	$\pm$	0.02	&	0.24	$\pm$	 0.01	&	76	\\ \hline
24	&	PCM	&	0.77	$\pm$	0.05	&	0.49	$\pm$	0.12	&	1.00	$\pm$	0.02	&	0.16	$\pm$	 0.01	&	7	\\ \hline
25	&	UKMO HADCM3	&	0.28	$\pm$	0.05	&	0.56	$\pm$	0.11	&	0.94	$\pm$	0.02	&	0.18	 $\pm$	0.01	&	42	\\ \hline
26	&	UKMO HADGEM1	&	0.52	$\pm$	0.04	&	0.63	$\pm$	0.10	&	1.05	$\pm$	0.02	&	0.20	 $\pm$	0.01	&	24	\\ \hline
	&	average	&	0.30	$\pm$	0.22	&	0.35	$\pm$	0.41	&	1.11	$\pm$	0.47	&	0.19	$\pm$	 0.11	&	143.8	\\ \hline
\end{tabular}
  \caption{Values of the regression parameters of Eq. \ref{eq2} obtained by fitting the 25 IPCC [2007] climate GCM ensemble-mean  estimates. \#1 refers to the ensemble average of the GISS ModelE depicted in Figure 1a; \#2-\#26 refers to the 25 IPCC GCMs. Pictures and analysis concerning all 95 records including each single GCM run are shown in Section 2 in the Supplement File that accompanies this paper.  The optimum value of these regression parameters should be $a=b=c=1$ as presented in the first raw that refers to the regression coefficients of the same model used to fit the temperature record. The last column refers to a reduced $\chi^2$ test based on three coefficients a, b and c: see Eq. \ref{eq233}. This determines the statistical compatibility of the regression coefficient measured for the GCM models and those observed in the temperature. It is always measured a reduced $\chi^2 \gg 1$ for three degrees of freedom, which indicates that the statistical compatibility of the GCMs with the observed 60-year, 20-year temperature cycles plus the secular trending is less than 0.1\%. These GCM regression values are depicted in Figure 4: the regression coefficients for each available GCM simulation are reported in the Supplement file. The $\chi^2$ test in the first line refers to the compatibility of the proposed model in Eq. 3 relative to the ideal case of $a=b=c=1$ that gives a reduced $\chi^2=0.21$ which imply that the statistical compatibility of Eq. 3 with the temperature cycles plus the secular trending is about   90\%.
  The fit has been implemented using the nonlinear least-squares (NLLS)Marquardt-Levenberg algorithm.}\label{}
\end{table}

\begin{table}
  \centering
\begin{tabular}{|c|l|c|c|c|c|}
  \hline
\#	&	model	&	s		(10.44-year)	&	l		(9.1-year)	&	$\chi^2$ (abcsl)	&	RMS ($^0C$)	\\ \hline
0	&	temperature	&	1.06	$\pm$	0.16	&	0.99	$\pm$	0.10	&	0.15	&	0.051	\\ \hline
1	&	GISS ModelE	&	0.30	$\pm$	0.11	&	0.40	$\pm$	0.07	&	61	&	0.107	\\ \hline
2	&	BCC CM1	&	0.53	$\pm$	0.11	&	0.49	$\pm$	0.07	&	70	&	0.105	\\ \hline
3	&	BCCR BCM2.0	&	-0.11	$\pm$	0.15	&	0.06	$\pm$	0.09	&	137	&	0.158	\\ \hline
4	&	CGCM3.1 (T47)	&	-0.47	$\pm$	0.09	&	0.06	$\pm$	0.06	&	479	&	0.212	\\ \hline
5	&	CGCM3.1 (T63)	&	0.39	$\pm$	0.15	&	-0.11	$\pm$	0.09	&	337	&	0.220	\\ \hline
6	&	CNRM CM3	&	0.22	$\pm$	0.24	&	-0.07	$\pm$	0.14	&	202	&	0.229	\\ \hline
7	&	CSIRO MK3.0	&	-0.54	$\pm$	0.14	&	-0.01	$\pm$	0.09	&	128	&	0.169	\\ \hline
8	&	CSIRO MK3.5	&	-0.53	$\pm$	0.13	&	0.44	$\pm$	0.08	&	134	&	0.156	\\ \hline
9	&	GFDL CM2.0 	&	-0.26	$\pm$	0.16	&	0.62	$\pm$	0.10	&	25	&	0.113	\\ \hline
10	&	GFDL CM2.1 	&	0.13	$\pm$	0.23	&	0.98	$\pm$	0.14	&	34	&	0.170	\\ \hline
11	&	GISS AOM	&	0.19	$\pm$	0.09	&	0.10	$\pm$	0.05	&	73	&	0.101	\\ \hline
12	&	GISS EH	&	0.27	$\pm$	0.14	&	0.66	$\pm$	0.09	&	30	&	0.106	\\ \hline
13	&	GISS ER	&	0.29	$\pm$	0.11	&	0.48	$\pm$	0.07	&	25	&	0.094	\\ \hline
14	&	FGOALS g1.0	&	-0.69	$\pm$	0.29	&	0.23	$\pm$	0.17	&	111	&	0.252	\\ \hline
15	&	INVG ECHAM4	&	-0.35	$\pm$	0.16	&	-0.23	$\pm$	0.10	&	105	&	0.132	\\ \hline
16	&	INM CM3.0	&	-0.15	$\pm$	0.24	&	1.01	$\pm$	0.14	&	36	&	0.150	\\ \hline
17	&	IPSL CM4	&	0.49	$\pm$	0.19	&	0.48	$\pm$	0.11	&	68	&	0.137	\\ \hline
18	&	MIROC3.2 Hires	&	0.17	$\pm$	0.16	&	0.43	$\pm$	0.09	&	69	&	0.122	\\ \hline
19	&	MIROC3.2 Medres	&	0.24	$\pm$	0.11	&	0.47	$\pm$	0.07	&	69	&	0.106	\\ \hline
20	&	ECHO G	&	0.52	$\pm$	0.13	&	0.54	$\pm$	0.08	&	20	&	0.097	\\ \hline
21	&	ECHAM5/MPI-OM	&	0.15	$\pm$	0.12	&	-0.09	$\pm$	0.07	&	82	&	0.126	\\ \hline
22	&	MRI CGCM 2.3.2	&	0.04	$\pm$	0.10	&	0.25	$\pm$	0.06	&	103	&	0.114	\\ \hline
23	&	CCSM3.0	&	0.12	$\pm$	0.13	&	0.91	$\pm$	0.08	&	50	&	0.110	\\ \hline
24	&	PCM	&	1.01	$\pm$	0.16	&	0.70	$\pm$	0.09	&	5	&	0.093	\\ \hline
25	&	UKMO HADCM3	&	0.07	$\pm$	0.15	&	-0.34	$\pm$	0.09	&	49	&	0.123	\\ \hline
26	&	UKMO HADGEM1	&	-0.46	$\pm$	0.14	&	0.32	$\pm$	0.08	&	30	&	0.107	\\ \hline
	&	average	&	0.06	$\pm$	0.40	&	0.34	$\pm$	0.37	&	97.39	&	0.139	\\ \hline
\end{tabular}
  \caption{Values of the regression parameters $s$ and $l$ of Eq. \ref{eq266} obtained by fitting the 26 IPCC [2007] climate GCM ensemble-mean  estimates. The fit has been implemented using the nonlinear least-squares (NLLS)Marquardt-Levenberg algorithm. Note that the two regression coefficients are  quite different from the optimum values $s=l=1$, as found for the temperature. The  column referring to the reduced $\chi^2$ test is based on all five regression coefficients (a, b, c, s and l) by extending  Eq. \ref{eq233}. Again it is always observed a $\chi^2\gg1$, which indicates incompatibility between the GCM and the temperature patterns.  The last column indicates the RMS residual values between the 4-year average smooth curves of each GCM simulation and the 4-year average  smooth curve of the temperature: the  value associated to the first raw (\emph{temperature})  RMS=0.051 $^oC$) refers to the RMS of the astronomical harmonic model that suggests that the latter is statistically 2-5 times more accurate than the GCM simulations in reconstructing the temperature record.}\label{}
\end{table}

\includepdf[pages={-}]{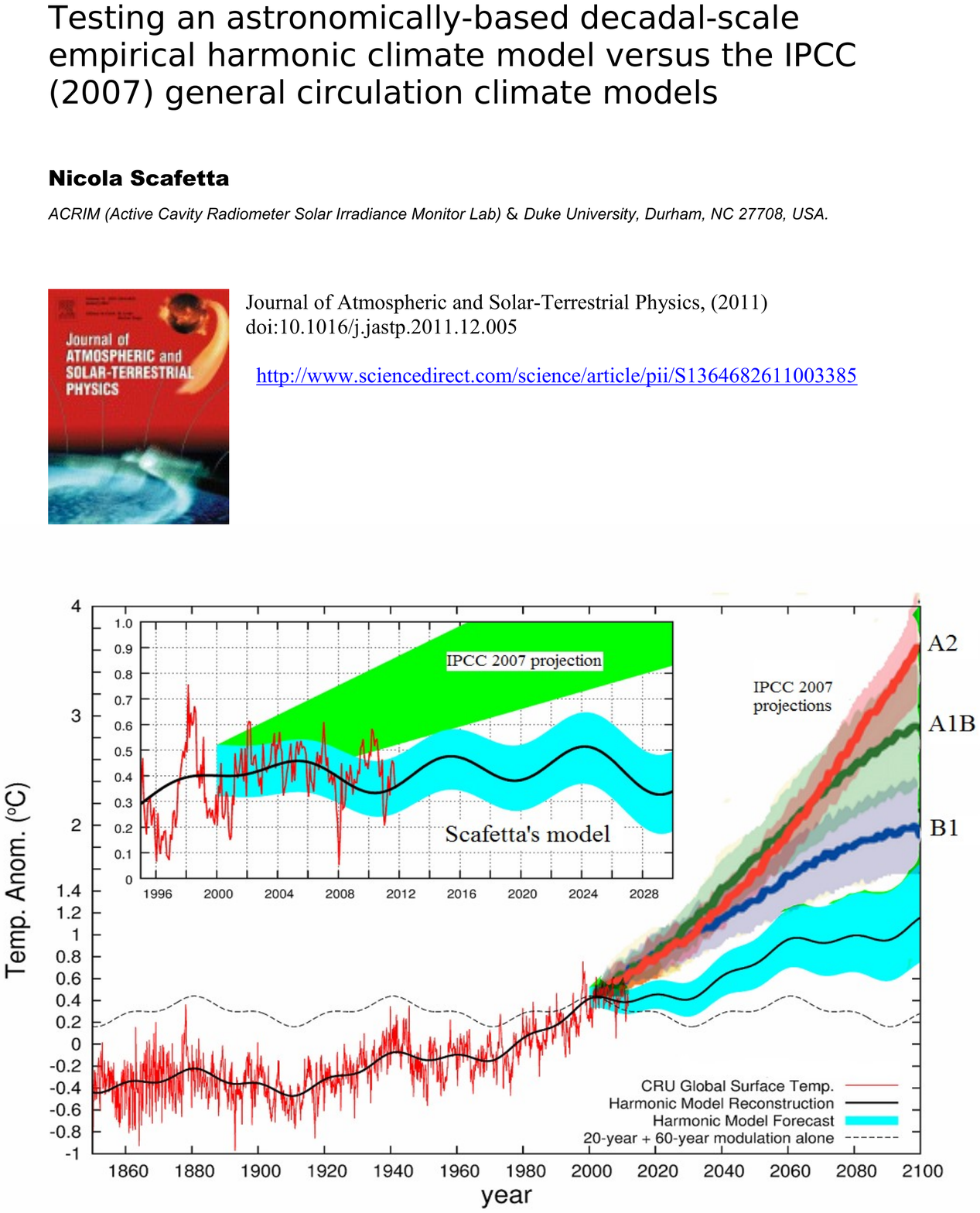}

\end{document}